\documentclass[aps,prl,twocolumn,showpacs,groupedaddress]{revtex4}  
\usepackage{graphicx}  
\usepackage{dcolumn}   
\usepackage{bm}        
\usepackage{amssymb}   
\usepackage{epsfig}

\begin{document}

\hspace{5.2in} \mbox{Fermilab-Pub-07/423-E}

\title{Search for flavor-changing-neutral-current $D$ meson decays}
%
\author{V.M.~Abazov$^{35}$}
\author{B.~Abbott$^{75}$}
\author{M.~Abolins$^{65}$}
\author{B.S.~Acharya$^{28}$}
\author{M.~Adams$^{51}$}
\author{T.~Adams$^{49}$}
\author{E.~Aguilo$^{5}$}
\author{S.H.~Ahn$^{30}$}
\author{M.~Ahsan$^{59}$}
\author{G.D.~Alexeev$^{35}$}
% ...

\affiliation{\vspace{0.1 in}(The D\O\ Collaboration)\vspace{0.1 in}}
% ...
\affiliation{$^{5}$University of Alberta, Edmonton, Alberta, Canada,
                Simon Fraser University, Burnaby, British Columbia, Canada,
                York University, Toronto, Ontario, Canada, and
                McGill University, Montreal, Quebec, Canada}
% ...
\affiliation{$^{28}$Tata Institute of Fundamental Research, Mumbai, India}
% ...
\affiliation{$^{30}$Korea Detector Laboratory, Korea University, Seoul, Korea}
% ...
\affiliation{$^{35}$Joint Institute for Nuclear Research, Dubna, Russia}
% ...
\affiliation{$^{49}$Florida State University, Tallahassee, Florida 32306, USA}
% ...
\affiliation{$^{51}$University of Illinois at Chicago,
                Chicago, Illinois 60607, USA}
% ...
\affiliation{$^{59}$Kansas State University, Manhattan, Kansas 66506, USA}
% ...
\affiliation{$^{65}$Michigan State University,
                East Lansing, Michigan 48824, USA}
% ...
\affiliation{$^{75}$University of Oklahoma, Norman, Oklahoma 73019, USA}
% ...
\date{August 15, 2007}

\begin{abstract}
We study the flavor-changing-neutral-current process $c\rightarrow u
\mu^+ \mu^-$ using $1.3$ fb$^{-1}$ of $p \bar{p}$
collisions at $\sqrt{s} = 1.96$ TeV recorded by the D0 detector
operating at the Fermilab Tevatron Collider.
We see clear indications of the $D_s^+$ and $D^+\rightarrow  \phi \pi^+ \rightarrow \mu^+\mu^-  \pi^+$ final states
with significance greater than four standard
deviations above background for the $D^+$ state. 
  We search for the continuum decay of $D^+\rightarrow \pi^+\mu^+\mu^-$ in the
dimuon invariant mass spectrum away from the $\phi$ resonance.  
We see no evidence of signal above background and set a limit
 of  ${\cal B}(D^+\rightarrow \pi^+\mu^+\mu^-) < 3.9\times 10^{-6}$  at the $90\%$ C.L. 
This limit places the most stringent constraint on new phenomena in the  $c\rightarrow u \mu^+ \mu^-$ transition.
\end{abstract}

\maketitle 


Many extensions of the standard model (SM) provide a mechanism for flavor-changing-neutral-current
 (FCNC) decays of beauty, charmed, and strange hadrons that could significantly alter
 the decay rate with respect to SM expectations.
  Since FCNC processes are forbidden at tree level in the SM, new physics effects could become visible
 in FCNC processes if the new amplitudes are larger than the higher-order penguin and box diagrams
 that mediate FCNC decays in the SM. In $B$ meson decays, the experimental sensitivity has reached the
 SM expected rates for many FCNC processes. In contrast, GIM suppression~\cite{bib:gim} in $D$ meson
 decays is significantly stronger and the SM branching fractions are expected
 to be as low as $10^{-9}$~\cite{bib:pakvasa,bib:fajer}.  
This leaves a large window of opportunity still available to search for new physics in charm decays.
There are several models of new phenomena such as SUSY $R$-parity 
violation in a single coupling scheme~\cite{bib:agashe} 
that lead to a tree level interaction mediated by new particles, or 
little Higgs models with a new up-like vector quark~\cite{bib:fajer2}
 that lead to direct $Z\rightarrow cu$ couplings. 
In both scenarios deviations from the SM might only be seen in the up-type quark sector, 
motivating the extension of 
experimental studies of FCNC processes to the charm sector.

 In this Letter we report on a study of FCNC charm decays including the
 first observation of the decay $D_s^+ \rightarrow \phi\pi^+ \rightarrow
  \mu^+ \mu^-  \pi^+$ and the first evidence for the decay $D^+ \rightarrow \phi\pi^+ \rightarrow
  \mu^+ \mu^- \pi^+$ by requiring a dimuon mass window
 around the nominal $\phi$ mass. The inclusion of charge conjugate modes is
  implied throughout the text.  
 At the reported level of statistics, we expect no contributions from two
 body $D^+_{(s)}$ decays due to the smaller $D^+_{(s)}\rightarrow \eta$, $\rho$, 
and $\omega$ branching fractions and the smaller $\eta$, $\rho$, 
and $\omega\rightarrow \mu^+\mu^-$ branching fractions~\cite{bib:pdg}. 
The search for the $c\rightarrow  u \mu^+\mu^-$ transition in the decay  
$D^+\rightarrow \pi^+\mu^+\mu^-$ is performed in the continuum region
 of the dimuon invariant mass spectrum below and above the $\phi$ resonance.
We focus on the $D^+$ continuum decay as opposed to similar $D_s^+$ or 
$\Lambda_c$ decays due to the longer lifetime and higher production fraction of the $D^+$ meson.
 The study uses a data sample of  $p\bar{p}$ collisions at
 $\sqrt{s}=1.96$ TeV corresponding to an integrated luminosity of  approximately $1.3$ fb$^{-1}$
recorded by the D0 detector operating at the
 Fermilab Tevatron Collider.
  Similar studies have recently been published by the FOCUS~\cite{bib:focus} and
 CLEO-c~\cite{bib:cleo} collaborations, and preliminary results have been presented by the BaBar~\cite{bib:babar} collaboration.

D0 is a general purpose detector described in detail in Ref.~\cite{bib:dzero,bib:dzeromuon}.
Charged particles are reconstructed using a silicon vertex tracker
 and a scintillating fiber tracker located inside a
superconducting solenoidal coil that provides a magnetic field of approximately $2$ T.
Photons and electrons are reconstructed using the inner region of a liquid argon
calorimeter optimized for electromagnetic shower detection. Jet
reconstruction and electron identification are further augmented with
the outer region of the calorimeter optimized for hadronic shower
detection. 
Muons are reconstructed using a
spectrometer consisting of magnetized iron toroids and three super-layers of
proportional tubes and plastic trigger scintillators located outside the calorimeter. 

The analysis is based on data collected with dimuon triggers.  
The D0 trigger is based on a three-tier system.  The level 1 and 2
dimuon triggers rely on hits in the muon spectrometer and
fast reconstruction of muon tracks.  The level 3 trigger performs fast
reconstruction of the entire event allowing for further muon
identification algorithms, matching of muon candidates to tracks
reconstructed in the central tracking system, and requirements on the
$z$ position of the $p\bar{p}$ interaction.

The selection requirements are determined using {\sc pythia}~\cite{bib:pythia} Monte
Carlo (MC) events to model both $c\bar{c}$ and $b\bar{b}$ production and
fragmentation.  The
{\sc evtgen}~\cite{bib:evtgen} MC is used to decay prompt $D$ mesons and secondary $D$ mesons
from $B$ meson decay into the $\phi \pi^+$ and $\mu^+\mu^-  \pi^+$
intermediate and final states. 
 The detector response is modeled using a {\sc geant}~\cite{bib:geant} based MC. 
The dimuon trigger is modeled using a detailed simulation program incorporating all aspects of the trigger logic.
  Backgrounds are
modeled using data in the mass sideband regions around the $D$ meson mass of $1.4 < m(\pi^+\mu^+\mu^-) < 1.6$ 
GeV$/c^2$ and $2.2 < m(\pi^+\mu^+\mu^-) < 2.4$ GeV$/c^2$.

\begin{figure}[htbp]
\centerline{\epsfysize 2.5 truein\epsfbox{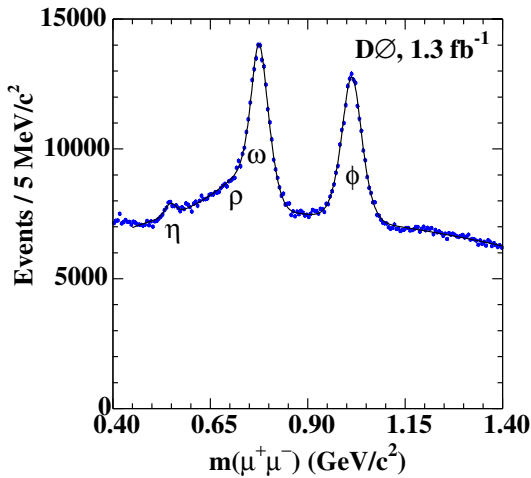}}
\caption{ The inclusive $m(\mu^+\mu^-)$ invariant mass spectrum. The fitting function includes components from 
the $\eta$, $\rho$, $\omega$, and $\phi$ resonances.}
\label{fig:mmumu}
\end{figure}

Muon candidates are required to have segments reconstructed in at
least two out of the three muon system super-layers and to be associated with a track
reconstructed with hits in both the silicon and fiber trackers. 
 We
require that the muon transverse momentum $p_T$ be greater than $2$ GeV$/c$ and
the total momentum $p$ to be above $3$ GeV$/c$.  
The dimuon system is formed
by combining two oppositely charged muon candidates that are associated
with the same track jet~\cite{bib:jet}, form a well
reconstructed vertex, and have an
invariant mass $m(\mu^+\mu^-)$ below $2$ GeV$/c^2$. 
The dimuon mass distribution
in the region of the light quark-antiquark resonances is shown
in Fig.~\ref{fig:mmumu}. Maxima corresponding to the
 production of $\omega$ and $\phi$ mesons are seen.  
The $\rho$ is observed as a broad structure beneath the 
$\omega$ peak, and there is some indication of $\eta$ production as well.  
For the initial search for resonance dimuon production we 
require the $\mu^+\mu^-$ mass be within $\pm 0.04$ GeV$/c^2$ of the nominal $\phi$ mass
and redetermine the muon momenta
with a $\phi$ mass constraint imposed~\cite{bib:pdg} which 
improves the $\mu^+\mu^-\pi^+$ invariant mass resolution by $33\%$.  

Candidate $D^+_{(s)}$
mesons are formed by combining the dimuon system with a track that is
associated with the same track jet as the dimuon
system, has
hits in both the silicon and fiber trackers, and has $p_T>0.18$ GeV$/c$. 
The pion impact parameter significance ${\cal S}_\pi$, defined as 
the point of closest approach of the track helix to the interaction point in the
 transverse plane relative to its error, is required to be greater than $0.5$.
 The
invariant mass of the three body system must be in the range $1.4$ GeV$/c^2$
$<m(\pi^+\mu^+\mu^-)<2.4$ GeV$/c^2$.   The three particles must form a well-reconstructed $D$ meson candidate 
vertex displaced from the primary vertex.  The transverse flight length significance ${\cal
  S}_D$, defined as the
transverse distance of the reconstructed $D$ vertex from the primary vertex
normalized to the error in the reconstructed flight length, is required to be greater than $5$.
The collinearity angle 
$\Theta_D$, defined as the angle between the $D$
momentum vector and the position vector pointing from the primary to
the secondary vertex, is required to be less than $500$ mrad.  In events with multiple $p\bar{p}$ collisions, the longitudinal track impact parameters are used to reject muons and tracks produced in the secondary $p\bar{p}$ interactions. 
In events with multiple $D$ candidates, the best candidate is chosen based
 on the $\chi^2_{\rm  vtx}$ of the three track vertex and the angular separation between the pion and the dimuon system in $\eta$-$\phi$ 
space, $(\Delta R_\pi)^2 = (\Delta \eta)^2 + (\Delta \phi)^2$, which is typically small for true candidates.
 
\begin{figure}[htbp]
\centerline{\epsfysize 2.5 truein\epsfbox{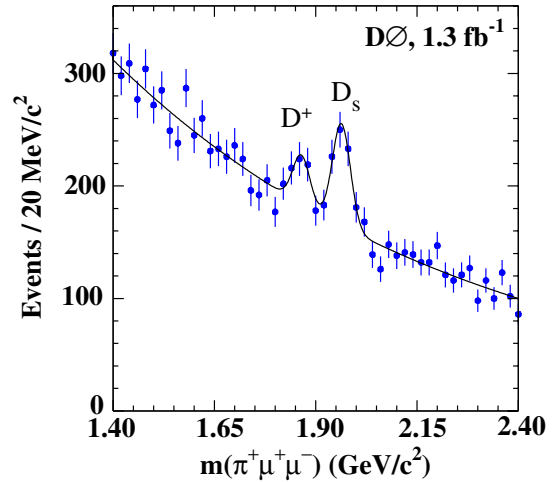}}
\caption{The $m(\pi^+\mu^+\mu^-)$ mass spectrum in the $0.98 < m(\mu^+\mu^-) < 1.06$ GeV$/c^2$ $\phi$ mass window.  The result of a binned likelihood fit to the distribution
  including contributions for $D^+$, $D^+_s$, and combinatoric
  background is overlaid on the histogram. }
\label{fig:loose}
\end{figure}

The resulting $\pi^+\mu^+\mu^-$
 invariant mass distribution is shown in Fig.~\ref{fig:loose}.
The $D^+_{(s)}\rightarrow \phi\pi^+ \rightarrow \mu^+\mu^- \pi^+$
 signal is extracted from a binned likelihood fit to
 the data
 assuming possible contributions from $D^+$ and
$D_s^+$ initial states as signal and from combinatoric background. 
The $D_s^+$ component is modeled by a Gaussian function with the mean and standard deviation as free parameters.  
The $D^+$ component is modeled as a Gaussian function. 
The difference in means between the $D^+$ and $D^+_s$ Gaussian functions is 
 constrained by the known mass difference and the ratio 
of the standard deviations is constrained to the ratio of masses~\cite{bib:pdg}. The background 
is modeled as an exponential function with floating parameters.
  The normalization of all functions are free parameters.
  The fit yields $254 \pm 36$ $D_s^+$
candidates and $115 \pm 31$ $D^+$ candidates.  
The statistical significance of the combined $D^+_s$ and $D^+$ 
signal is $8$ standard deviations above background. The significance 
of the $D^+$ yield, treating both the combinatorial and $D^+_s$ 
candidates as background, is $4.1$ standard deviations.

The relative efficiency of the $D^+$ and $D_s^+$ channels is 
determined separately for prompt $D$ mesons produced in 
direct $p\bar{p}\rightarrow c\bar{c} + X$ processes and $D$ mesons from $B$ meson decay and 
combined using the measured prompt fractions~\cite{bib:cdfD} 
$\epsilon^+ =  f^+_p\epsilon^+_{\rm prompt} + (1-f^+_p)\epsilon^+_{B\rightarrow D}$,
where $\epsilon^+_{\rm prompt}$ is the efficiency for prompt $D^+$ mesons, $\epsilon^+_{B\rightarrow D}$
 is the efficiency for $D^+$ mesons from $B$ meson decay, 
and $f^+_p$ is the fraction of prompt $D^+$ mesons; we use equivalent expressions for $D_s^+$ mesons.
The yield ratio is related to the ratio of branching fractions by
$${n(D^+) \over n(D^+_s)} = {f^+_{c\rightarrow D} \over f^s_{c\rightarrow D}}  {f^s_p \over f^+_p}  {\epsilon^+ \over \epsilon^s}  {{\cal B}(D^+\rightarrow \phi\pi^+ \rightarrow \mu^+\mu^-  \pi^+) \over {\cal B}(D^+_s\rightarrow \phi\pi^+)  {\cal B}(\phi \rightarrow \mu^+\mu^-)},$$
where $f^+_{c\rightarrow D}$ is the fraction of $D^+$ mesons produced in $c$ quark 
fragmentation, and $f^s_{c\rightarrow D}$ is the equivalent fraction for $D_s^+$ mesons~\cite{bib:zeus}.  
The relevant numbers are listed in Table~\ref{tab:inputs}.  
The efficiency ratio is determined from MC to be $\epsilon^s/\epsilon^+ = 0.70 \pm 0.06$ (stat + sys). 
 The difference from unity is caused by the lifetime difference between $D^+_s$ ($c\tau = 147.0$ $\mu$m)
 and $D^+$ ($c\tau = 311.8$ $\mu$m) mesons, and the systematic uncertainty is dominated by uncertainties
 in the resolution modeling of ${\cal S}_D$ and ${\cal S}_\pi$.

\begin{table}
\caption{\label{tab:inputs} External inputs to the yield ratio calculation. All numbers have been corrected for the most recent $D^+\rightarrow K^+\pi^+\pi^-$ and $D_s^+\rightarrow \phi\pi^+$ branching fractions~\cite{bib:pdg}.}
\begin{ruledtabular}
\begin{tabular}{lc}
$f^+_p$~\cite{bib:cdfD} & $0.891 \pm 0.004$\\ 
$f^s_p$~\cite{bib:cdfD} & $0.773 \pm 0.038$\\ 
\hline
$f^s_{c\rightarrow D}/f^+_{c\rightarrow D}$~\cite{bib:zeus} & $0.40 \pm 0.09$ \\ 
\end{tabular}
\end{ruledtabular}
\end{table}

Using the efficiency ratio, production fractions, and the $D_s^+\rightarrow\phi\pi^+$ and $\phi\rightarrow \mu^+\mu^-$ branching fractions gives
$ {\cal B}(D^+\rightarrow \phi\pi^+ \rightarrow \mu^+\mu^-  \pi^+) =  (1.8 \pm 0.5 \hskip 1 mm {\rm ( stat.)}\hskip 1mm \pm 0.6 \hskip 1mm {\rm (sys.)}) \times 10^{-6}$,
which is consistent with the expected value of $(1.86 \pm 0.26) \times 10^{-6}$ given by 
the product of the $D^+ \rightarrow \phi\pi^+$ and $\phi\rightarrow \mu^+\mu^-$ branching
 fractions and other recent measurements~\cite{bib:cleo,bib:babar}.  The systematic uncertainty 
is overwhelmingly dominated by the uncertainty in the $D_s^+\rightarrow \phi\pi^+$ branching 
fraction that enters both the normalization and $f^s_{c\rightarrow D}$. 

We now turn to the search for the continuum decay of $D^+\rightarrow \pi^+\mu^+\mu^-$ mediated by FCNC interactions.
We study the dimuon invariant mass
 region below $1.8$ GeV$/c^2$, excluding $0.96 < m(\mu^+\mu^-) < 1.06$ GeV$/c^2$. 
Backgrounds are further reduced by requirements on the ${\cal
  S}_D$,  ${\cal S}_\pi$, $\Theta_D$,  $\chi^2_{\rm  vtx}$, and $\Delta R_\pi$ variables 
defined above. The pion transverse momentum $p_T(\pi)$ and the
 isolation defined as 
${\cal I}_D = p(D) /\sum{p_{\rm cone}}$
where the sum is over tracks in a cone centered on
the $D$ meson of radius $\Delta R = 1$ are also used.  
The final requirements are chosen using a random grid search~\cite{bib:random}
 optimized using the Punzi~\cite{bib:punzi} criteria to give the optimal $90\%$ C.L. upper limit.  
The final requirements along with the number of candidates surviving each requirement are listed in Table~\ref{tab:cuts}.

\begin{table}
\caption{\label{tab:cuts} Final requirements for the background suppression variables along with the number of candidates surviving each requirement for the continuum  $D^+\rightarrow \pi^+\mu^+\mu^-$ analysis.}
\begin{ruledtabular}
\begin{tabular}{lcc}
 & Requirement & Surviving candidates \\
\hline
Preselection & & $154203$ \\
$\Delta R_\pi$      & $<2.6$ & $154131$ \\
$p_T(\pi)$      & $>0.4$ GeV/$c$ & $127027$ \\
${\cal S}_D$    & $>9.4$     & $69817$ \\
${\cal S}_\pi$  & $>1.8$     & $51736$ \\
${\cal I}_D$    & $>0.7$     &   $24742$ \\
$\Theta_D$      & $<7$ mrad  & $962$ \\
$\chi^2_{\rm vtx}$ (3 DOF)  & $<2.6$     & $212$ \\
Signal window  & $\pm 2\sigma$     & $19$ \\

\end{tabular}
\end{ruledtabular}
\end{table}

\begin{figure}[htbp]
\centerline{\epsfysize 2.5 truein\epsfbox{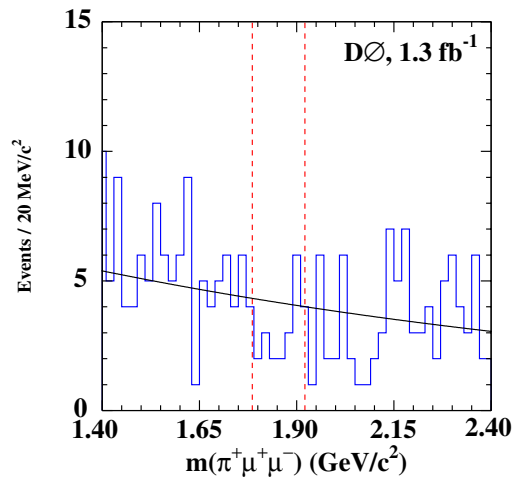}}
\caption{Final $\pi^+\mu^+\mu^-$ invariant mass spectrum.  The $\pm 2\sigma$ $D^+$ signal region, within the dashed lines, contains $19$ events.  The background level determined from the sidebands is $25.8 \pm 4.6$ events.}
\label{fig:pimumu-final}
\end{figure}

The $\pi^+\mu^+\mu^-$ invariant mass distribution in data for the dimuon invariant mass
 region below $1.8$ GeV$/c^2$ excluding $0.96 < m(\mu^+\mu^-) < 1.06$ GeV$/c^2$ is 
shown in Fig.~\ref{fig:pimumu-final}.
The $D^+$ signal region contains 19 events. The combinatorial background in the signal 
region is estimated by performing sideband extrapolations to be $25.8 \pm 4.6$ events.
 The uncertainty reflects the range in the background estimation from variation in the 
background shape across the  $\pi^+\mu^+\mu^-$ mass spectrum.  
The probability of the background fluctuating to the measured event yield or fewer events is $14\%$.

We normalize the results to the $D^+\rightarrow \phi\pi^+ \rightarrow \mu^+\mu^-  \pi^+$ signal 
instead of the larger $D_s^+$ signal to avoid the uncertainties associated with the $D^+$ and 
$D^+_s$ production fractions.  We use the product of the known $D^+\rightarrow \phi\pi^+$ and 
$\phi\rightarrow \mu^+\mu^-$ branching fractions~\cite{bib:pdg}.  
The signal efficiency ratio between
 the  $D^+\rightarrow \pi^+\mu^+\mu^-$ channel in the final sample and the $D^+\rightarrow \phi\pi^+\rightarrow \mu^+\mu^-  \pi^+$ channel in the preselection sample is determined from MC to be $(5.4 \pm 0.8)\%$.  
The inputs to the limit
 calculation are summarized in Table~\ref{tab:limits}.
The systematic uncertainty is dominated by the modeling of the vertex resolution particularly in the $\chi^2_{\rm vtx}$ requirement.
  Using this, we find
$${ {\cal B}(D^+\rightarrow \pi^+\mu^+\mu^-) \over {\cal B}(D^+\rightarrow \phi\pi^+)\times {\cal B}(\phi \rightarrow \mu^+\mu^-)} < 2.09, \hskip 2mm 90\% \hskip 1mm \rm C.L.$$
The limit is determined using a Bayesian technique~\cite{bib:limit}.
Using the central value of $D^+\rightarrow\phi\pi^+$ and $\phi\rightarrow \mu^+\mu^-$ 
branching fractions gives
$$ {\cal B}(D^+ \rightarrow \pi^+\mu^+\mu^-) < 3.9 \times 10^{-6},  \hskip 2mm 90\% \hskip 1mm \rm C.L.$$
This is approximately $30\% $ below the limit one would expect to set given 
an expected background of $25.8\pm 4.6$ events.  
The single event sensitivity, given by the branching fraction one would derive based
 on one observed signal candidate, is $3.0\times 10^{-7}$.

\begin{table}
\caption{\label{tab:limits} Inputs to the  ${\cal B}(D^+ \rightarrow \pi^+\mu^+\mu^-)$ upper limit calculation and resulting upper limit at the $90\%$ and $95\%$ C.L.}
\begin{ruledtabular}
\begin{tabular}{lc}
$D^+ \rightarrow \pi^+\mu^+\mu^-$ candidate yield & $19$ events \\
Background expectation & $25.8\pm 4.6$ events \\
$D^+ \rightarrow \phi\pi^+\rightarrow \pi^+\mu^+\mu^-$ candidate yield & $115 \pm 31$ events \\
Relative efficiency & $0.054\pm 0.008$ \\
${\cal B}(D^+ \rightarrow \phi \pi^+)$ & $6.50\times 10^{-3}$ \\
${\cal B}(\phi \rightarrow \mu^+\mu^-)$ & $2.86\times 10^{-4}$ \\
Single event sensitivity & $3.0\times 10^{-7}$ \\
\hline
${\cal B}(D^+ \rightarrow \pi^+\mu^+\mu^-)$ $95\%$ C.L. & $<6.1 \times 10^{-6}$ \\
${\cal B}(D^+ \rightarrow \pi^+\mu^+\mu^-)$ $90\%$ C.L. & $<3.9 \times 10^{-6}$ \\
\end{tabular}
\end{ruledtabular}
\end{table}

In conclusion, we have performed a detailed study of $D^+$ and $D_s^+$ 
decays to the $\pi^+\mu^+\mu^-$ final state.
  We clearly observe the $D_s^+\rightarrow \phi\pi^+$ intermediate state 
and see evidence for the $D^+\rightarrow \phi\pi^+$
 intermediate state.  The branching fraction for the 
$D^+\rightarrow \phi\pi^+ \rightarrow \pi^+\mu^+\mu^-$ final
 state is consistent with the product of $D^+\rightarrow 
\phi \pi^+$ and $\phi\rightarrow \mu^+\mu^-$ branching fractions.
  We have performed a search for the continuum decay
 of $D^+\rightarrow \pi^+\mu^+\mu^-$ by excluding the region of
 the dimuon invariant mass spectrum around the $\phi$.
  We see no evidence of signal above background and set a limit
 of $ {\cal B}(D^+ \rightarrow \pi^+\mu^+\mu^-) < 3.9 \times 10^{-6} $ 
at the $90\%$ C.L. This is the most stringent limit to date in 
a decay mediated by a $c\rightarrow u\mu^+\mu^-$ transition. Although this is approximately 
 500 times above the SM expected rate, it already reduces the 
allowed parameter space of the product of SUSY $R$-parity violating 
couplings $\lambda^\prime_{22k}\times \lambda^\prime_{21k}$~\cite{bib:pakvasa}.  However, it is still an order 
of magnitude above the expected level from little Higgs models~\cite{bib:fajer2}.

%
We thank the staffs at Fermilab and collaborating institutions.  We also thank Sandip Pakvasa for several useful discussions.  We acknowledge support from the 
DOE and NSF (USA);
CEA and CNRS/IN2P3 (France);
FASI, Rosatom and RFBR (Russia);
CAPES, CNPq, FAPERJ, FAPESP and FUNDUNESP (Brazil);
DAE and DST (India);
Colciencias (Colombia);
CONACyT (Mexico);
KRF and KOSEF (Korea);
CONICET and UBACyT (Argentina);
FOM (The Netherlands);
Science and Technology Facilities Council (United Kingdom);
MSMT and GACR (Czech Republic);
CRC Program, CFI, NSERC and WestGrid Project (Canada);
BMBF and DFG (Germany);
SFI (Ireland);
The Swedish Research Council (Sweden);
CAS and CNSF (China);
Alexander von Humboldt Foundation;
and the Marie Curie Program.
%

\end{document}